\documentclass[12pt]{amsart}
\usepackage{latexsym, amssymb, enumerate, amsmath}
\usepackage{graphics}
\usepackage{algpseudocode}
\usepackage{graphicx}
\usepackage{hyperref}

\newcommand{\half}{\frac12}

\newcommand{\R}{\mathbb{R}}

\title{Estimating differential entropy using recursive copula splitting}

\author{Gil Ariel}
\address{Department of Mathematics, Bar Ilan University, Ramat Gan, Israel}
\email{arielg@math.biu.ac.il}
\author{Yoram Louzoun}
\email{louzouy@math.biu.ac.il}
\begin{document}

\bibliographystyle{unsrt}

\begin{abstract}
A method for estimating the Shannon differential entropy of multidimensional random variables using independent samples
is described.
The method is based on decomposing the distribution into a product of the marginal distributions and the joint dependency,
also known as the copula.
The entropy of marginals is estimated using one-dimensional methods.
The entropy of the copula, which always has a compact support,  
is estimated recursively by splitting the data along statistically dependent dimensions.
The method can be applied both for distributions with compact and non-compact supports,
which is imperative when the support is not known or of mixed type (in different dimensions).
At high dimensions (larger than 20), numerical examples demonstrate that our method is not only more accurate, 
but also significantly more efficient than existing approaches.
\end{abstract}

\maketitle

%%%%%%%%%%%%%%%%%%%%%%%%%%%%%%%%%%%%%%%%%%%%%%%%%%%%%%%%%
\section{Introduction}
\label{sec:intro}
%%%%%%%%%%%%%%%%%%%%%%%%%%%%%%%%%%%%%%%%%%%%%%%%%%%%%%%%%

\setcounter{equation}{0}

Differential entropy (DE) is used in a range of fields and disciplines, including signal processing and
machine learning, evaluation of independence \cite{Calsaverini2009} and feature selection \cite{Kwak2002,Kerroum2010,Zhu2010}.
The problem is also related to dimension reduction through independent component analysis \cite{Faivishevsky2009}
and quantifying order in out-of-equilibrium physical systems \cite{Avinery2017,Martiniani2019}.
The DE of a multi-dimensional distribution with density $p(x):\R^D \to \R$ is defined as,
\begin{equation}
       H = - \int_{\R^n} p(x) \ln p(x) dx .
\label{eq:DE}
\end{equation}
Despite a large number of suggested algorithms \cite{Beirlant1997review,Paninski2003}, 
the problem of estimating the DE from independent sampling
of the distributions remains a challenge in high dimensions.
Broadly speaking, algorithms can be classified as one of two approaches: binning and sample-spacing methods,
or their multidimensional analogues - partitioning and nearest-neighbor (NN) methods .
In 1D, the most straight-forward method is to partition the support of the distribution into bins
and either calculate the entropy of the histogram or use it for plug-in estimates \cite{Granger1994,Beirlant1997review,Sricharan2010}.
This amounts to approximating $p(x)$ as a piece-wise constant function (i.e., assuming that the distribution
is uniform in each subset in the partition).
This works well if the support of the underlying distribution is bounded and given.
If the support is not known or is unbounded, it can be estimated as well,
for example using the minimal and maximal observations.
In such cases, sample-spacing methods \cite{Beirlant1997review} that use the spacings between adjacent samples are advantageous.
Overall, the literature provides a good arsenal of tools for estimating 1D entropy including rigorous bounds on
convergence rates (given some further assumptions of $p$). See \cite{Beirlant1997review,Paninski2003} for reviews.

Estimating entropy in higher dimensions is significantly more challenging \cite{Darbellay1999,Paninski2003}.
Binning methods become impractical as having $M$ bins in each dimension implies $M^D$ bins overall. Beyond the computational costs, most such bins will often have 1 or 0 samples, leading to a significant underestimate of the entropy.
In order to overcome this difficulty, Stowell and Plumbley \cite{Stowell2009} suggested partitioning the data using a $k$-D
partitioning tree-hierarchy ($k$DP).
In each level of the tree, the data is divided into two parts with an equal number of samples.
The splitting continues recursively across the different dimensions (see below for a discussion on the stopping criteria).
The construction essentially partitions the support of $p$ into bins that are multi-dimensional rectangles whose sides are aligned with the principal axes.
The DE is then calculated assuming a uniform distribution in each rectangle.
As shown below, this strategy works well at low dimensions (typically 2-3) and only if the support is known.
The method is highly efficient, as constructing the partition tree has an $O(N \log N)$ efficiency.
In particular, it has no explicit dependence on the dimension.

Spacing methods are generalized using the set of $k$ nearest neighbors to each sample ($k$NN) \cite{Kozachenko1987,Kraskov2004,Sricharan2010,Gao2013,Lord2018}.
These are used to locally approximate the density, typically using kernels \cite{Joe1989,Granger1994,Singh2003,Shwartz2005,Ozertem2008,Gao2016}.
As shown below, $k$NN schemes preform well at moderately high dimensions (up to 10-15)
for distributions with unbounded support.
However, they fail completely when $p$ has a compact support and become increasingly inefficient with the dimension.
Broadly speaking, algorithms for approximating $k$NN in $D$-dimensions have an efficiency of $\epsilon^{-D} N \log N$, where $\epsilon$ is
the required accuracy \cite{Indyk2016}.
Other approaches for entropy estimation include variations and improvements of $k$NN (e.g. \cite{Singh2003,Faivishevsky2009,Gao2016}), Voronoi-based partitions \cite{Miller2003} (which are also prohibitively expensive at very high dimensions), Parzen windows \cite{Kwak2002} and ensemble estimators \cite{Sricharan2013}.

Here, we follow the approach of Stowell and Plumbley \cite{Stowell2009}, partitioning space using trees.
However, we add an important modification that significantly enhances the accuracy of the method.
The main idea is to decompose the density $p(x)$ into a product of marginal (1D) densities and a copula. The copula is computed over the compact support of the one dimensional cumulative distributions.  As such, the multidimensional DE estimates become the combination of one dimensional estimates, and a multi-dimensional estimate on a compact support, even if the support of the original distribution was not compact. We term the proposed method as CopulA Decomposition Entropy Estimate (CADEE).

Following Sklar's theorem \cite{Book:copula,Durante2010}, any continuous multi-dimensional density $p(x)$ can be written uniquely as
\begin{equation}
       p(x) = p_1(x_1) \cdot \dots \cdot p_D (x_D) c(F_1(x_1),\dots,F_D(x_D)) ,
\label{eq:Sklar}
\end{equation}
where, $x=(x_1,\dots,x_D)$, $p_k(\cdot)$ denotes the marginal density of the $k$'th dimension 
with cumulative distribution function (CDF) $F_k (t) = \int_{-\infty}^t p_k(x) dx$,
and $c(u_1,\dots,u_D)$ is the density of the copula, i.e., a probability density on the hyper-square $[0,1]^D$ whose
marginals are all uniform on $[0,1]$,
\begin{equation}
       \left[ \Pi_{j=1 \dots D, ~j \neq k} \int du_j \right] c(u_1, \dots , u_D) = 1,
\label{eq:uniformMarginals}
\end{equation}
for all $k$. 
Substituting \eqref{eq:Sklar}  into \eqref{eq:DE} yields,
\begin{equation}
       H = \sum_{k=1}^D H_k + H_c,
\label{eq:Centropy}
\end{equation}
where $H_k$ is the entropy of the $k$'th marginal, to be computed using appropriate 1D estimators,
and $H_c$ is the entropy of the copula.
Using Sklar's theorem has been previously suggested as a method for calculating the mutual information between variables,
which is identical to the copula entropy $H_c$ \cite{Calsaverini2009,Giraudo2013,Hao2015,Xue2017}.
The new approach here is in showing that $H_c$ can be efficiently estimated recursively, similar to the $k$DP approach.

Splitting the overall estimation into the marginal and copula contributions has several major advantages.
First, the support of the copula is compact, which is exactly the premise for which partitioning methods are most adequate.
Second, since the entropy of the copula is non-positive, adding up the marginal entropies across tree-levels
provides an improving approximation (from above) of the entropy. 
Finally, the decomposition brings-forth a natural criterion for terminating the tree-partitioning and for dimension reduction
using pairwise independence.

The following sections are organized as follows. Section~\ref{sec:method} describes the outline of the CADEE algorithm.
In order to demonstrate its wide applicability, several examples in which the DE can be calculated analytically are presented.
In addition, our results are compared to previously suggested methods.
Section~\ref{sec:implementation} discusses implementation issues and the algorithm's computational cost.
We conclude in section~\ref{sec:summary}.

%%%%%%%%%%%%%%%%%%%%%%%%%%%%%%%%%%%%%%%%%%%%%%%%%%%%%%%%%
\section{CADEE method}
\label{sec:method}
%%%%%%%%%%%%%%%%%%%%%%%%%%%%%%%%%%%%%%%%%%%%%%%%%%%%%%%%%

The main idea proposed here is to write the entropy $H$ as a sum of $D$ 1D marginal entropies,
and the entropy of the copula.
Analytically, the copula is obtained by a change of variables,
\begin{equation}
       c (u_1,\dots,u_D) = p ( F_1 (x_1) , \dots , F_D (x_D) ) ,
\label{eq:cFromP}
\end{equation}
Let $x^i = (x^i_1, \dots , x^i_D) \in \R^D$, $i=1 \dots N$ denote $N$ independent samples from a real $D$-dimensional
random variable (RV) with density $p(x)$.
We would like to use the samples $x^i$ in order to obtain samples from the copula density $c(u_1, \dots, c_D)$.
From \eqref{eq:cFromP}, this can be obtained by finding the rank (in increasing order) of samples along each dimension.
In the following, this operation will be referred to as a rank transformation.
This is the empirical analogue of the integral transform where one plugs the sample into the CDF.
More formally, for each $k=1 \dots D$, let $\sigma_k$ denote a permutation of $\{1 \dots N \}$ that arranges $x^1_k,\dots x^N_k$ in increasing order, i.e.,
$x^{\sigma_k^i}_k \le x^{\sigma_k^j}_k$ for $i \le j$.
Then, taking
\begin{equation}
       u_k^i = \frac{1}{N} (\sigma_k^i - 1/2) ,
%\label{eq:Centropy}
\end{equation}
yields $N$ samples $u^i = (u^i_1, \dots , u^i_D) \in [0,1]^D$, $i=1 \dots N$ from the distribution $c (u_1,\dots,u_D)$.
Note that the samples are not independent.
In other words, the rank is the emperical CDF, shifted by $1/2N$.
In particular, they correspond to $N$ distinct points on a uniform grid, $u^i \in \{ 1/2N, 3/2N , 1-1/2N \}^D$.
%Also note that the samples are not mapped to $\{0,1/(N-1), \dots ,1 \}$ because that would effectively change the support of the copula at subsequent iterations.

\begin{figure}[h!tp]
\centering
\includegraphics[trim={3cm 5.5cm 3cm 0cm},clip, width=9cm, angle=0]{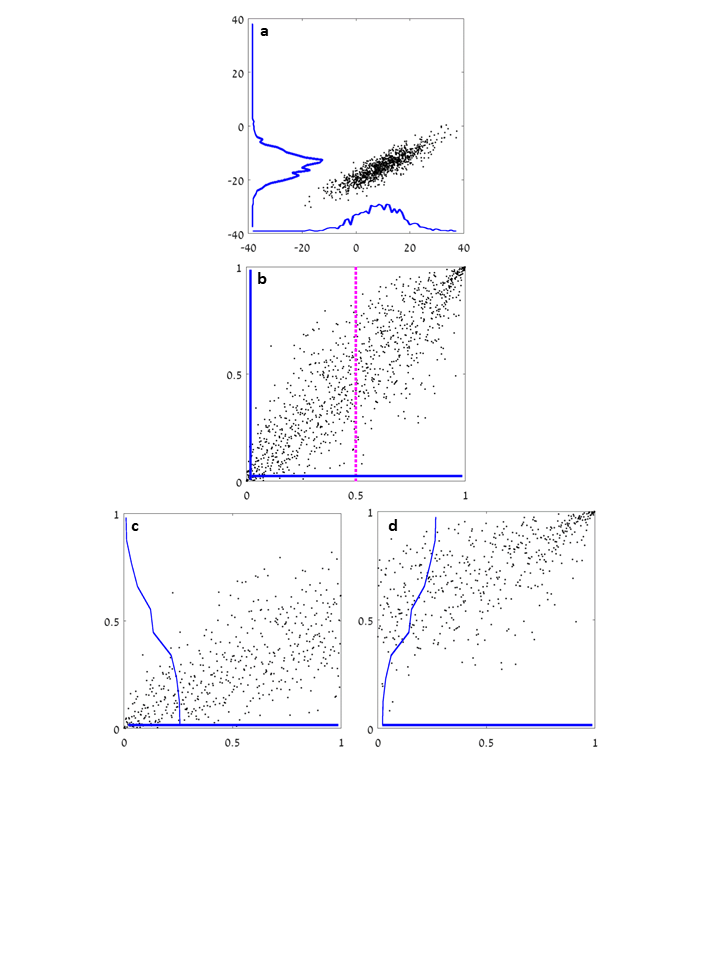}
\caption{
 A schematic sketch of the proposed method.
 {\bf a.} A sample of 1000 points from a 2D Gaussian distribution. The blue lines depict the empirical density (obtained using uniform bins).
 {\bf b.} Following the rank transform (numbering the sorted data in each dimension), the same data provides samples for the copula in $[0,1]^2$.
 Splitting the data according to the median in one of the axes (always at 0.5) yields {\bf c} (left half) and {\bf d} (right half).
 The blue lines depict the empirical density in each half.
 Continue recursively.
}
\label{fig:sketch}
\end{figure}

1D entropies are estimated using either uniform binning or sample-spacing methods, depending on whether the support of the
marginal is known to be compact (bins) or unbounded/unknown (spacing).
The main challenge lies in evaluating the DE of high-dimensional copulas \cite{Calsaverini2009,Embrechts2013}.
In order to overcome this difficulty, we compute it recursively, following the $k$DP approach.
Let $k \in \{1,\dots,D \}$ be a spatial dimensions, to be chosen using any given order.
The copula samples $u^i$ are split into two equal parts (note that the median in each dimension is $1/2$).
Denote the two halves as $v_j^i = \{ u_j^i | u_k^i \le 1/2 \}$ and $w_j^i = \{ u_j^i | u_k^i > 1/2 \}$.
Scaling the halves as $2v_j^i$ and $2w_j^i-1$ produces two sample sets for two new copulas, each with $N/2$ points.
A simple calculation shows that
\begin{equation}
       H_c = \frac12 ( H_{2v} + H_{2w-1} ) ,
%\label{eq:Centropy}
\end{equation}
where $H_{2v}$ is the entropy estimate obtained using the set of points $2v_j^i$
and $H_{2w-1}$ is the entropy estimate obtained using the set of points $2w_j^i-1$.
The marginals of each half may no longer be uniformly distributed in $[0,1]$, which suggests
continuing recursively, i.e., the entropy of each half is a decomposed using Sklar's them, etc.
See Fig.~\ref{fig:sketch} for a schematic sketch of the method.

A key question is finding a stopping condition for the recursion.
In \cite{Stowell2009}, Stowell and Plumbley apply a statistical test for uniformity of $x_k$, the dimension used for splitting.
This condition is meaningless for our method as copulas have uniform marginals by construction.
In fact, this suggests that one reason for the relatively poor $k$DP estimates at high $D$ is the rather simplistic stopping criterion,
requiring that only one of the marginals is statistically similar to a uniform RV.

In principle, we would like to stop the recursion once the copula cannot be statistically distinguished from
the uniform distribution on $[0,1]^D$.
However, reliable statistical tests for uniformity at high $D$ are essentially equivalent to evaluating the copula entropy \cite{Joe1989,Calsaverini2009,Embrechts2013}.
As a result, we relax the stopping condition to only test for pairwise dependence. The precise test for that will be further discussed.
Calculating pairwise dependencies also allows a dimension reduction approach:
if the matrix of pairwise-dependent dimensions can be split into blocks, then each block can be treated independently.

\begin{figure}[h!tp]
\centering
\includegraphics[trim={1cm 10cm 2cm 1cm},clip, width=12cm, angle=0]{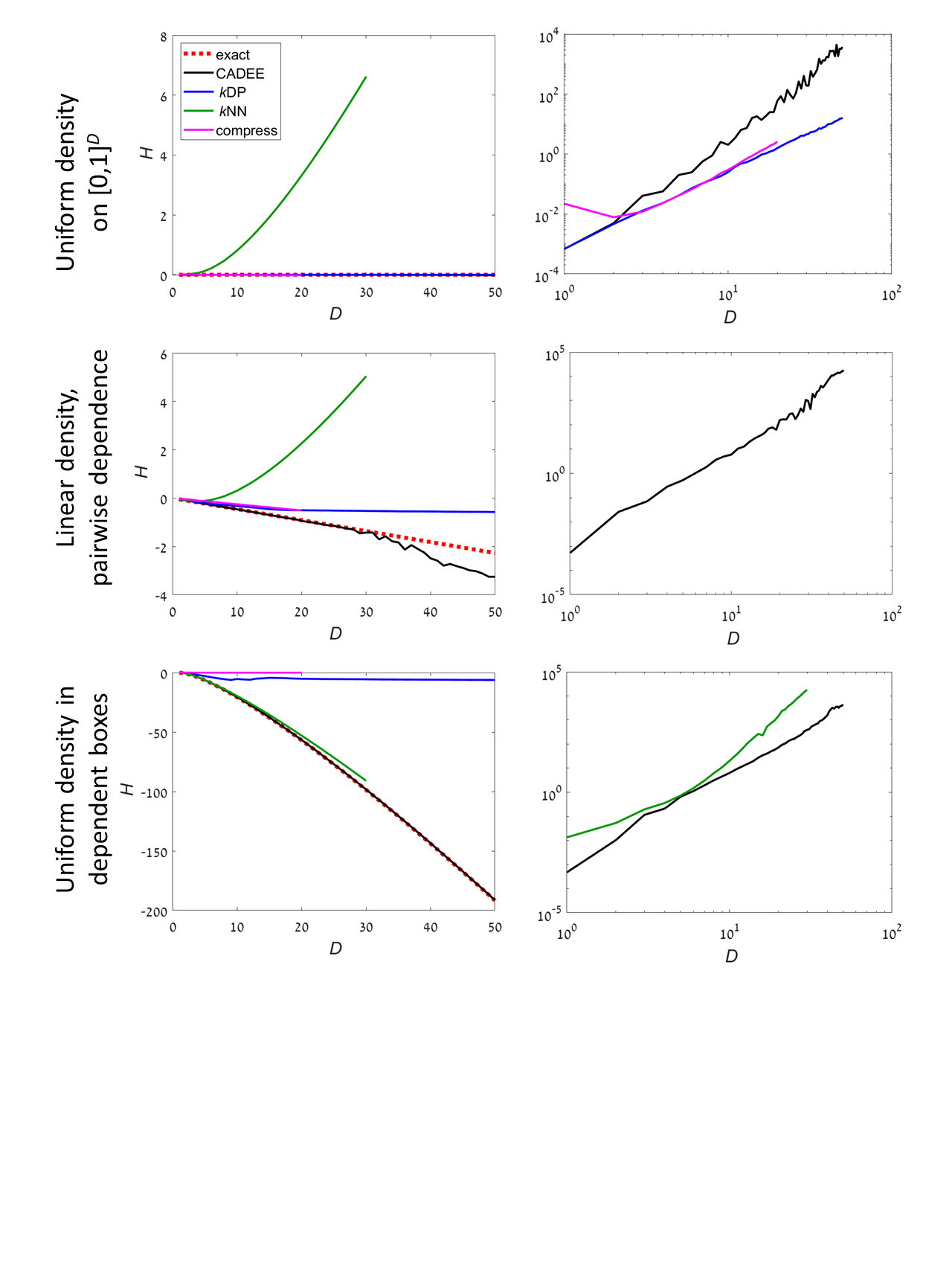}
\caption{
  Estimating the entropy for given analytically-computable examples (dashed red line) with compact distributions ($[0,1]^D$). Black: using the recursive copula splitting method, Blue: $k$DP, Green: $k$NN, magenta: lossless compression (magenta).
  Left: The estimated entropy as a function of dimension. Right: Running times (on a log-log scale), showing only relevant methods.
  The number of samples is $N=10,000D^2$.
  See also Tables~\ref{tbl:results10} and \ref{tbl:results20} for detailed numerical results with $D=10$ and 20.
}
\label{fig:resultsC}
\end{figure}

\begin{figure}[h!tp]
\centering
\includegraphics[trim={1cm 19cm 2cm 1cm},clip, width=12cm, angle=0]{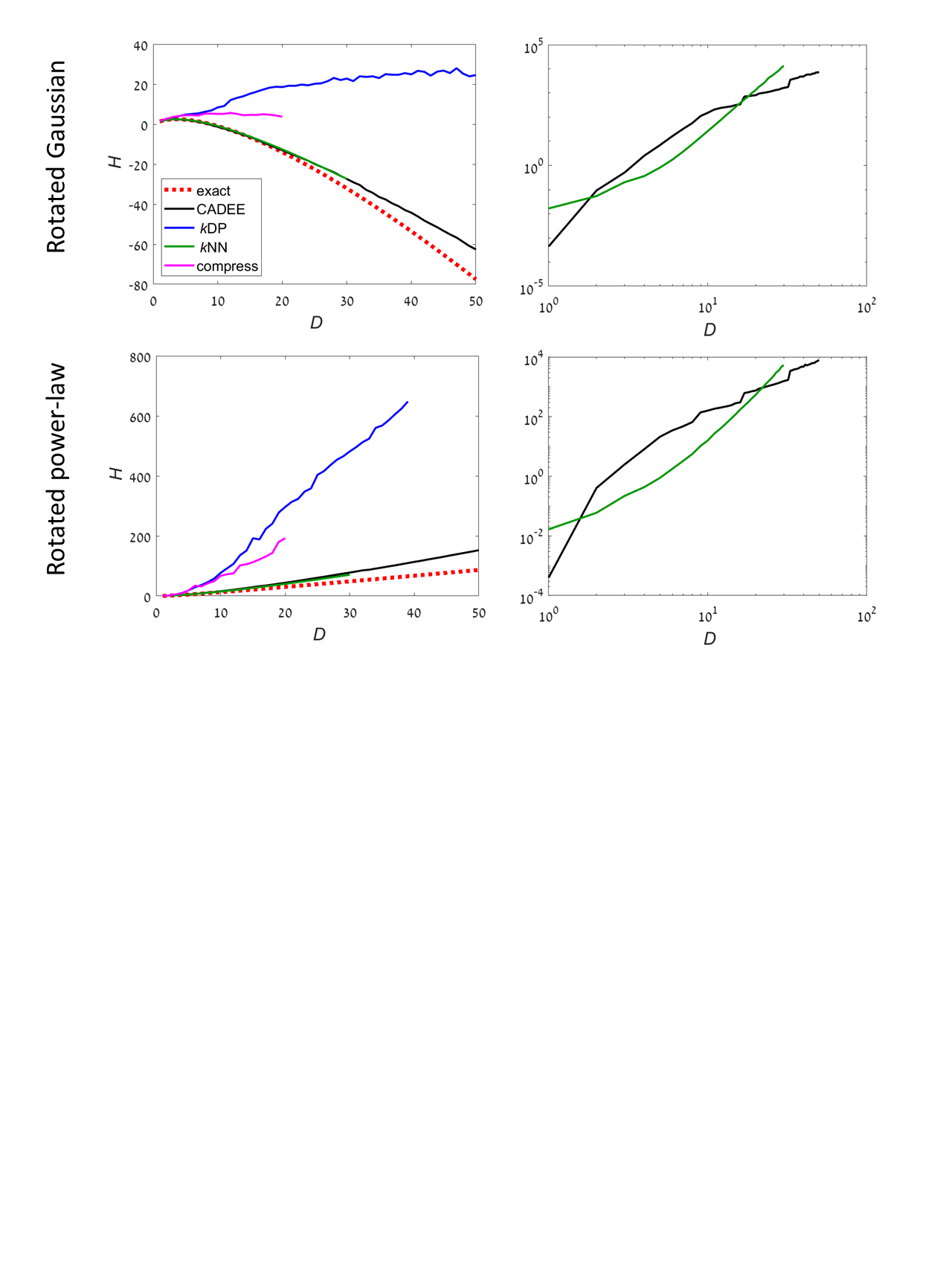}
\caption{
  Estimating the entropy for given analytically-computable examples (dashed red line) with non-compact distributions. Black: using the recursive copula splitting method, Blue: $k$DP, Green: $k$NN, magenta: lossless compression (magenta).
  Left: The estimated entropy as a function of dimension. Right: Running times (on a log-log scale), showing only relevant methods.
    The number of samples is $N=10,000D^2$.
    The inaccuracy of our and the $k$NN method is primarily due to the relatively small number of samples.
    See also Tables~\ref{tbl:results10} and \ref{tbl:results20} for detailed numerical results with $D=10$ and 20.
}
\label{fig:resultsNC}
\end{figure}

In order to demonstrate the applicability of the method described above, we study the results of our algorithm for several distributions
for which the DE \eqref{eq:DE} can be computed analytically.
Figures~\ref{fig:resultsC} and \ref{fig:resultsNC} show numerical results for $H$ and the running time as a function of
dimension using an implementation in Matlab.
Five different distributions are studied.
Three have a compact support in $[0,1]^D$ (Fig.~\ref{fig:resultsC}):
\begin{itemize}
\item {\bf C1}: A uniform distribution.
\item {\bf C2}: Dependent pairs. The dimensions are divided into pairs. The density in each pair is $p(x,y)=x+y$, supported on $[0,1]^2$.
Different pairs are independent.
\item {\bf C3}: Independent boxes. Uniform density in a set consisting of $D$ small hypercubes, $\cup_{k=1}^D [(k-1)/D,k/D]^D$.
\end{itemize}
Two examples have an unbounded support (Fig.~\ref{fig:resultsNC}):
\begin{itemize}
\item {\bf UB1}: Gaussian distribution. The covariance is chosen to be a randomly rotated diagonal matrix with eigenvalues $k^{-2}$, $k=1 \dots D$.
Then, the samples are rotated to a random orthonormal basis in $\R^D$.
The support of the distribution is $\R^D$.
\item {\bf UB2}: Power-law distribution. Each dimension $k$ is sampled independently from a density $x^{-2-2/k}$, $k=1 \dots D$ in $[1,\infty)$.
Then, the samples are rotated to a random orthonormal basis in $\R^D$.
The support of the distribution is a $2^{-D}$ fraction of $\R^D$ that is not aligned with the principal axes.
\end{itemize}
Results with our method are compared to three algorithms:
\begin{enumerate}
\item The $k$DP algorithm \cite{Shwartz2005}. We use the C implementation available in \cite{GitHubKDP}.
\item The $k$NN algorithm based on the Kozachenko-Leonenko estimator \cite{Kozachenko1987}. We use the C implementation available in \cite{GitHubKL}.
\item A lossless compression approach \cite{Avinery2017,Martiniani2019}. Following \cite{Avinery2017}, samples are binned into 256 equal bins in each dimension, and the data is converted into an $N\times D$ matrix of 8-bit unsigned integers.
The matrix is compressed using the LZW algorithm (implemented in Matlab's imwrite function to a gif file).
In order to estimate the entropy, the file size is interpolated linearly between a constant matrix (minimal entropy)
and a random matrix with independent uniformly distributed values (maximal entropy), both of the same size.
\end{enumerate}
Theoretically, in order to get rigorous convergence of estimators, the number of samples should grow exponentially with
the dimension \cite{Beirlant1997review}.
Since this requirement is impractical at very high dimensions, we consider an under-sampled case and only use
$N=10,000 D^2$ samples.
Each method was tested at increasing dimensions until a running time of about 3 hours was reached (per run, on a standard PC) or
the implementation ran out of memory. In such cases, no results are reported for this and following dimensions.
See also Tables~\ref{tbl:results10} and \ref{tbl:results20} for numerical results for $D=10$ and 20.

Note that, in principle, it may be advantageous to apply a PCA of SVD of the sample convariance to decouple dependent directions.
Such methods will be particular advantageous for the unbounded problems. 
We do not apply such conventional pre-processing methods here in order to make it {\it more difficult} for the CADEE method.
If SVD converges the distribution into a product of independent 1D variables, the copula is close to 1 and the method
will be highly exact after a single iteration.

For compact distributions, it is well known than $k$NN methods may fail completely.
This can be seen even for the most simple examples such as uniform distributions (example C1).
However, $k$NN works well in example C3 because the density occupies a small fraction of the volume,
which is optimal for $KNN$.
$k$DP and compression methods are precise for uniform distribution, which is a reference case for these methods.
For examples C2 and C3, both are highly inaccurate at $D>5$.
In comparison, CADEE shows very good accuracy up to $D=30-50$, depending on the example.

For unbounded distributions, $k$DP and compression methods do not provide meaningful results for $D>3$.
Both CADEE and $k$NN provide good estimates up to $D=20$ ($k$NN is slightly better), but diverge slowly at higher dimensions (CADEE is better).
Numerical tests suggest this is primarily due to the relatively small number of samples, which severely under-samples the distributions at high $D$.
Comparing running times, the recursive copula splitting method is significantly more efficient at high dimensions.
Simulations suggest a polynomial running time (see Section~\ref{sec:implementation} for details),
while $k$NN is exponential in $D$, becoming prohibitively inefficient at $D>30$.

%%%%%%%%%%%%%%%%%%%%%%%%%%%%%%%%%%%%%%%%%%%%%%%%%%%%%%%%%
\section{Convergence analysis}
\label{sec:convergence}
%%%%%%%%%%%%%%%%%%%%%%%%%%%%%%%%%%%%%%%%%%%%%%%%%%%%%%%%%

In this section, we study the convergence properties of CADEE, i.e., the estimation error as $N$ increases with fixed $D$.
We proceed along three routes. 
First, we consider an example in which the first several copula splittings can be
preformed analytically. 
The example demonstrates how, ignoring statistical errors, recursive splitting of the copula and adding up the marginal entropies at the different recursion levels gets close to the exact entropy. Next, we provide a general analytical bound on the error of the model.
Although the bound is not tight, it establishes that, in principle, the method provides a valid approximation of the entropy.
Finally, we study the convergence of the method numerically for several low dimensional examples, providing empirical evidence that
the rate of convergence of the method (the average absolute value of the error) is $O(N^{-\alpha})$ for some $0<\alpha<0.5$.

\subsection{Analytical example}

In order to demonstrate the main idea why splitting the copula iteratively improves the entropy estimate, we work-out 
a simple example in which the splittings can be performed analytically. 
For the purpose of this example, sampling errors in the estimate of the 1D entropy are neglected.

Consider the dependent pairs example (C2) with $D=2$. The two dimensional density of the sampled random variable is given by
\begin{equation}
   p(x,y)=x+y ,
\end{equation}
with support in $[0,1]^2$.
The exact entropy is $H= - \int_0^1 \int_0^1 p \ln dx dy = 5/6-4/3*\ln 2 \simeq -0.09086$.
In order to obtain the copula, we first write the marginal densities and CDFs,
\begin{equation}
\begin{aligned}
   p_X (x)=x+ \half,  ~~~ F_X(x) = \half x^2 + \half x \\
   p_Y (y)=y+ \half,  ~~~ F_Y(y) = \half y^2 + \half y 
\end{aligned}
\end{equation}
Using Sklar's theorem,
\begin{equation}
   p(x,y)=p_X(x) p_Y(y) c(F_x(x),F_Y(y)) .
\end{equation}
Since the CDFs are invertible (in $[0,1]$), it can be equivalently written as
\begin{equation}
   p(F_X^{-1}(s),F_Y^{-1}(t))=p_X(F_X^{-1}(s)) p_Y(F_X^{-1}(t)) c(s,t) .
\end{equation}
Hence, 
\begin{equation}
   c(x,y) = \frac{ -1 + \sqrt{1/4+2x} + \sqrt{1/4+2y} }{ \sqrt{1/4+2x}\sqrt{1/4+2y} } .
\end{equation}
Indeed, one verifies that the marginals are uniform,
\begin{equation}
   \int_0^1 c(x,y) dx= \int_0^1 c(x,y) dy = 1.
\end{equation}

Continuing the CADEE algorithm, we compute the entropy of marginals,
$H_X = H_Y = 1/2 - 9/8*\ln 3 + \ln 2 \simeq -0.04279$.
This implies that the copula entropy is $H-H_X-H_Y \simeq -0.00528$ (5.8\% of $H$).
In order to approximate it, we split $c(x,y)$ into two halves, for example along the $Y$ axis. 
Each density is shifted and stretched linearly to have support in $[0,1]^2$ again,
\begin{equation}
\begin{aligned}
   c_1(x,y) = \frac{ -1 + \sqrt{1/4+2x} + \sqrt{1/4+y} }{ \sqrt{1/4+2x}\sqrt{1/4+y} } \\
   c_2(x,y) = \frac{ -1 + \sqrt{1/4+2x} + \sqrt{5/4+y} }{ \sqrt{1/4+2x}\sqrt{5/4+y} } .
\end{aligned}
\end{equation}
We continue recursively, computing the marginals for $c_1$ and $c_2$
\begin{equation}
\begin{aligned}
   c_{1X}(x) = \sqrt{5}-1 + \frac{ 2-\sqrt{5} }{ \sqrt{1/4+2x}}, ~~~ c_{1Y}=1 \\
      c_{2X}(x) = 3-\sqrt{5} + \frac{ \sqrt{5} -2 }{ \sqrt{1/4+2x}}, ~~~ c_{2Y}=1 
\end{aligned}
\end{equation}
The marginal entropies are $H_{1X} = -0.00284$, $H_{1Y}=0$, $H_{1X} = -0.00267$ and $H_{2Y}=0$.
Overall, summing up the marginal entropies of the two iterations we have 
$H_X+H_Y+0.5 ( H_{1X}+H_{1Y}+H_{2X}+H_{2Y}) =-0.08834$ (error=2.77\%).

We continue similarly, calculating the copula of $c_1$ and $c_2$ and then the marginal distributions of their copulas.
We find that the entropy after the third iteration is $H_X+H_Y+0.5 ( H_{1X}+H_{1Y}+H_{2X}+H_{2Y} + 0.5 (H_{11X} +H_{11Y}+H_{12X} +H_{12Y}+H_{21X} +H_{21Y}+H_{22X} +H_{22Y}) ) = 0.08993$ (error= 1.02\%).

Indeed, we see that in the absence of statistical errors, the recursive splitting provides in improving upper bound for the entropy.

\subsection{Analytical bound}

Here, we provide an analytical estimate of the bias and statistical error incurred by the algorithm. 
We derive a bound, which is not tight. Detailed analysis of the bias and error in some adequate norm is beyond the scope of the current paper.

The first part of the analysis estimated the worst-case accuracy by iteratively approximating the entropy using $q$ repeated splittings of the copula.
In the last iterations, the dimensions are assumed to be independent, i.e., the copula equals 1.

%Using Sklar's theorem, any continuously differentiable density $p(x)$, $x \in \R^D$ can be written as
%
%\begin{equation}
%   p(x) = \Pi_{i=1}^D p_i(x_i) c (F_1 (x_1), \dots,F_D (x_D)).
%\end{equation}
%
%The total entropy is given by $H=\sum_i H_i + H_c$, where $H_i$ is entropy of $p_i$ and $H_c$ is the entropy of the copula.
Consider the copula $c(u_1,\dots,u_D)$, which is split, e.g. along $u_1 \in [0,1]$ into two halves corresponding to $u_1 \in [0,1/2]$ and $u_1 \in [1/2,1]$.
Linearly scaling back into $[0,1]$, we obtain two densities
\begin{equation}
\begin{aligned}
 &  c_1 (s,u_2,\dots,u_D) = c(s/2,u_2,\dots,u_D)  \\
 & c_2 (s,u_2,\dots,u_D) = c(1/2+s/2,u_2,\dots,u_D) ,
\end{aligned}
\end{equation}
where $(s,u_2,\dots,u_D) \in [0,1]^D$.
It is easily seen that $H_c = (H_{1c} + H_{2c})/2$, where $H_{1c}$ and $H_{1c}$ are the entropies of $c_1$ and $c_2$, respectively.
We continue recursively, splitting the resulting copulas along some dimension.
After $q$ iterations, we obtain and expression of the form,
\begin{equation}
   H = \sum_{j=1}^D \left[ H_i + \sum_{k=1}^q \frac{1}{2^k} \sum_{i_1,\dots,\i_k\in \{1,2\}} H_{i_1,\dots,i_k,j} \right] + 
      \frac{1}{2^q} \sum_{i_1,\dots,\i_k\in \{1,2\}} H_{i_1,\dots,i_k,c} ,
\label{eq:recursiveH}
\end{equation}
where $H_{i_1,\dots,i_k,j}$ is the 1D entropy of the $j$'th marginal and $H_{i_1,\dots,i_k,c}$ is the entropy of the copula, obtained after $k$ splittings along the dimensions $i_1,\dots,\i_k$.
For simplicity, we assume that the dimensions are chosen sequentially and suppose that $q=D^r$, i.e., each dimension was split $r$ times.

Let $\Delta=2^{-r}$ and 
suppose that the copula $c(x)$ is constant on small hyper-rectangles with sides 
\begin{equation}
[F_1^{-1}(i_1\Delta),F_1^{-1}((i_1+1)\Delta)]  \times \dots \times [F_D^{-1}(i_D\Delta),F_D^{-1}((i_D+1)\Delta)],
\end{equation}
where $i_k \in \{0,\dots, r-1\}$.
This implies that within these rectangles all dimensions are independent.
Then, $H_{i_1,\dots,i_D,c}=0$ and the last sum in \eqref{eq:recursiveH} vanishes.

Next, we approximate $c(x)$ in each small rectangle using Taylor. 
Without loss of generality, we focus on the case $i_1 = \dots = i_D = 0$.
To first order,
$c(x) = (A + B_1 x_1 + \dots B_D x_D)$, with $A,B_1,\dots,B_D$ are $O(1)$. 
Scaling to $[0,1]^D$, 
$c_\Delta (x) = Z^{-1} (A + B_1 F_1^{-1}(\Delta) x_1 + \dots B_D F_D^{-1}(\Delta) x_D)$,
where $Z$ is a normalization constant.
Assuming that $F_k$ are continuously differentiable and strictly increasing, 
$F_K^{-1}$ are also continuously differentiable and $F_k^{-1}(\Delta) = O(\Delta)$.
Then, since the total mass in each rectangle is exactly $\Delta^D$,
we have that $A/Z = 1 + O(\Delta)$.
Finally, the entropy of the normalized density  $c_\Delta (x)$ can be estimated.
Expanding the log to order 1 in $\Delta$,
\begin{equation}
   H [c_\Delta] = - \int_0^1 dx_1 \dots \int_0^1 dx_D c_\Delta (x) \ln c_\Delta (x)  = - D \ln ( 1/\Delta) + O(\Delta).
\end{equation}
From this, one needs to subtract $D \ln \Delta$ to compensate for the scaling.
Therefore, for any continuously differential, strictly positive (in its support) density, $H_{i_1,\dots,i_k,c} = O(\Delta)$.
We conclude that the entire last sum in \eqref{eq:recursiveH} sums to order $\Delta$.
The prefactor is typically proportional to $D$.

Next, we consider statistical errors. 
Using the Kolmogorov-Smirnov statistics, the distance between the empirical CDF and the exact one is of order $N^{-1/2}$.
Suppose 1D entropy estimates use a method with accuracy (absolute error) of order $N^{-\alpha}$, $\alpha \le 1/2$.
Then, in the worst case, if all errors are additive, then each estimate in the $k$'th iterate has an error (in absolute value) of order $(N/2^k)^{-\alpha}$.
Overall, we have
\begin{equation}
\begin{aligned}
   \Delta H = D \sum_{k=1}^q \frac{1}{2^k} \sum_{i_1,\dots,\i_k\in \{1,2\}} \left( \frac{N}{2^k} \right)^{-\alpha}
   = D \sum_{k=1}^q \left( \frac{N}{2^k} \right)^{-\alpha} \\
   = D N^{-\alpha}  \sum_{k=1}^q (2^\alpha)^k 
   \le   D N^{-\alpha}  ( \sum_{k=1}^q 2^k )^\alpha = D (2^{q+1}/N)^\alpha .
\end{aligned}
\end{equation}
For fixed $q$, the statistical error decreases like  $N^{-\alpha}$.
Typically, for an unbiased 1D estimator in which the variance of the estimator is of order $N^{-2\alpha}$,
the variance of the overall estimation using CADEE is
\begin{equation}
   {\rm Var} [\Delta H ]=  D (2^{q+1}/N)^{2\alpha}.
\end{equation}

However, the prefactor depends linearly on the dimension $D$ and exponentially on the number of iterations $q$.
Recall that the bias decreases exponentially with $q/D$.
Hence, the two sources of errors should be balanced in order to obtain a convergent approximation.

\subsection{Numerical examples}

\begin{figure}[h!tp]
\centering
\includegraphics[trim={1cm 5cm 1cm 3cm},clip, width=13cm, angle=0]{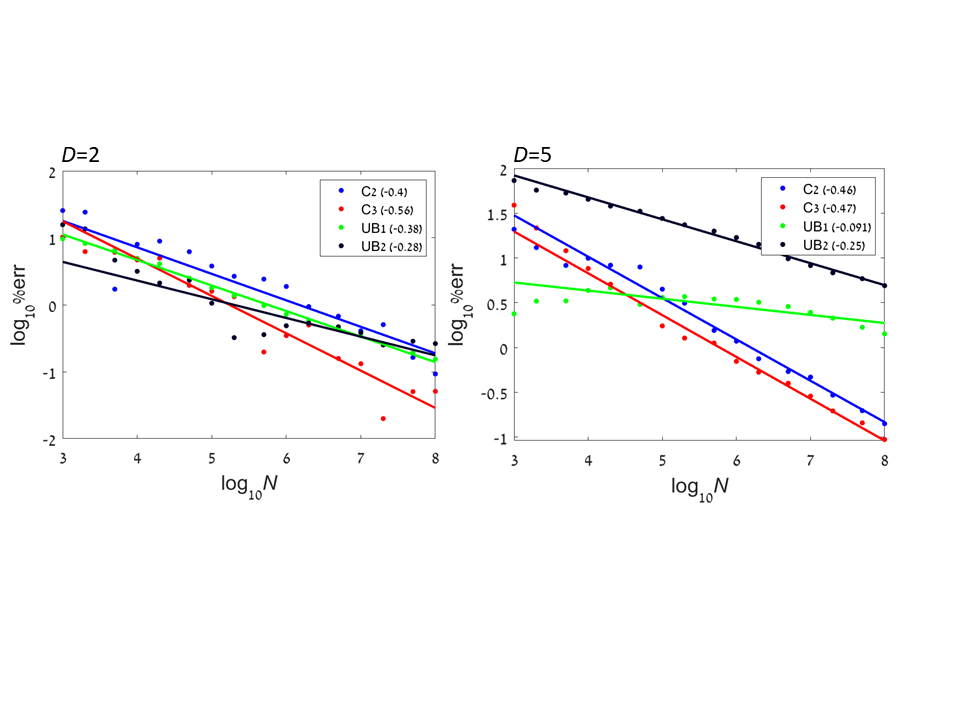}
\caption{
Convergence rates of CADEE: The average absolute value of the error as a function of $N$.
Left: $D=2$. Right $D=5$.
}
\label{fig:varyN}
\end{figure}

In order to demonstrate the convergence of the method, we test the error of the estimate obtained using CADEE for small $D$ examples.
Figure~\ref{fig:varyN} shows numerical results with four types of distributions (dependent pairs, independent boxes, Gaussian and power-law)
with $D=2$ and $D=5$ and $10^3$-$10^8$ samples.
As discussed above, larger dimensions require significantly more samples in order to guaranty that the entire support is sampled
at appropriate frequencies.
We see that for all examples, the method indeed converges. 
For non-bounded distributions, the rate decreases with dimension.

%%%%%%%%%%%%%%%%%%%%%%%%%%%%%%%%%%%%%%%%%%%%%%%%%%%%%%%%%
\section{Implementation details}
\label{sec:implementation}
%%%%%%%%%%%%%%%%%%%%%%%%%%%%%%%%%%%%%%%%%%%%%%%%%%%%%%%%%

The following is a pseudo-code implementation of the algorithm described above.
Several aspects of the codes, such as choice of constants, stopping criterion and
estimation of pair-wise independence are rather heuristic approaches, which were found to improve the
accuracy and efficiency of our method.
Recall that for every $i$, $(x_1^i,\dots,x_D^i) \in \R^D$ is an independent sample.
\begin{algorithmic}[H]
  \Function{copulaH}{$\{ x^i_k \}$, $D$, $N$, $level=0$}
    \State $H \gets 0$
    \For{$k$ = 1 to $D$}
    \State $u_k \gets$ rank($x_k$)/N \Comment{Calculate rank (by sorting)}
    \State $H \gets H + $ H1D($\{ u^i_k \}$, $N$, $level$) \Comment{entropy of marginal $k$}
    \EndFor\\
    \If{$D=1$ or $N<=$ min \#samples}
    \State {\Return $H$}
    \EndIf
    \\
    \State \Comment{$A$ is the matrix of pairwise independence}
    \State $A_{ij} =$ true if $x^i$ and $x^j$ are statistically independent 
    \State $n_{\rm blocks} \gets$ \# of blocks in $A$.
    \If{ $n_{\rm blocks}>1$} \Comment{Split dimensions}
    \For{$j=1$ to $n_{\rm blocks}$}
    \State $v \gets$ elements in block $j$
    \State $H \gets H + $ copulaH($\{ u_k^i \}_{k \in v}^{i=1\dots N}$,dim($v$),$N$,$level$)
    \EndFor
    \State \Return $H$
    \Else \Comment{No independent blocks}
    \State $k \gets$ choose a dim for splitting
    \State $L=\{ i | u_k^i \le 1/2 \}$
    \State $\{ v_j^i \} = \{ 2 u_j^i | i \in L, j=1 \dots D\}$
    \State $H \gets H +$ copulaH($\{ v_j^i \}$,$D$,$N/2$,$level+1$) /2
    \\
    \State $R=\{ i | u_k^i > 1/2 \}$
    \State $\{ w_j^i \} = \{ 2 u_j^i -1 | i \in R, j=1 \dots D \}$
    \State $H \gets H +$ copulaH($\{ w_j^i \}$,$D$,$N/2$,$level+1$) /2
    \EndIf
\EndFunction
\end{algorithmic}

Several steps in the above algorithm should be addressed.
\begin{enumerate}
  \item The rank of an array $x$ is the order in which values appear. Since the support of all marginals in the copula is $[0,1]$, we take ${\rm rank} (x) =\{ 1/2 , 3/2 , N-1/2\}$. For example, ${\rm rank} ([-2,0,-3]) =\{ 3/2 , 5/2 , 1/2\} $.
This implies that the minimal and maximal samples are not mapped into $\{0,1\}$, which would artificially change the support of the distribution.
  The rank transformation is easily done using sorting.
  \item 1D entropy: One-dimensional entropy of compact distributions (whose support is $[0,1]$) is estimated using a histogram with uniformly spaced bins.
  The number of bins can be taken to depend of $N$,
  and order $N^{1/3}$ is typically used (we used $N^{1/3}$ or $N^0.4$ for spacing or bin-based methods, respectively. 
  For additional considerations and methods for choosing the number of bins see \cite{Knuth2006}.
  At the first iteration, the distribution may not be compact, and the entropy is estimated using $m_N$-spacings (see \cite{Beirlant1997review}, Eq. (16)).
  \item Finding blocks in the adjacency matrix $A$:
  Let $A$ be a matrix whose entries are 0 and 1, where $A_{kl}=1$ implies that $u^k$ and $u^l$ are independent.
  By construction, $A$ is symmetric.
  Let $D$ denote the diagonal matrix whose diagonal elements are the sums of rows of $A$.
  Then, $L=A-D$ is the Laplacian associated with the graph described by $A$.
  In particular, the sum of all rows of $L$ is zero.
  We seek a rational basis for the kernel of a matrix $L$: Let ker($L$) denote the kernel of a matrix $L$.
  By a rational basis we mean an orthogonal basis (for ker($L$)), in which all the coordinates are either 0 or 1
  and the number of 1's is minimal.
  In each vector in the basis, components with 1's form a cluster (or block), which is pair-wise independent of all other marginals.
  In Matlab, this can be obtained using the command null($L$,'r').
  For example, consider the adjacency matrix 
  \[
  A = \left( \begin{tabular}{ccc} 1 & 0 & 1 \\ 0 & 1 & 0 \\ 1 & 0 & 1 \end{tabular} \right) ,
  \]
  whose graph Laplacian is 
  \[
  D = \left( \begin{tabular}{ccc} 2 & 0 & 0 \\ 0 & 1 & 0 \\ 0 & 0 & 2 \end{tabular} \right) ~,~~~ L = A - D = \left( \begin{tabular}{ccc} -1 & 0 & 1 \\ 0 & 0 & 0 \\ 1 & 0 & -1 \end{tabular} \right) ,
  \]
  A rational basis for the kernel of $L$ (which is 2D) is
  \[
  \left\{ \left( \begin{tabular}{c} 1 \\ 0 \\ 1 \end{tabular} \right) , \left( \begin{tabular}{c} 0 \\ 1 \\ 0 \end{tabular} \right) \right\} ,
  \]
\end{enumerate}
which corresponds to two blocks - components 1+3 and component 2.

Pairwise independence is determined as follows.
\begin{enumerate}
   \item Calculate the Spearman correlation matrix of the samples $\{ x_k\}$, denoted $R$.
   Note that this is the same as the Pearson correlation matrix of the ranked data $\{ u_k\}$.
  \item Assuming normality and independence (which does not hold), the distribution of elements in $R$ is asymptotically given by the t-distribution with $N-2$ degrees of freedom.
  Denoting the CDF of the t-distribution with $n$ degrees of freedom by $T_{n} (z)$, two marginals $(k,l)$ are considered uncorrelated if $|R_{kl}|>T^{-1}_{n-2} (1-\alpha/2)$, where $\alpha$
  is the acceptance threshold.
  We take the standard $\alpha=0.05$.
  Note that because we do $D(D-1)/2$ tests, the probability of observing independent vectors by chance grows with $D$.
  This can be corrected by looking at the statistics of the maximal value for $R$ (in absolute value), which tends to a Gumbel distribution \cite{Han2017}.
This approach (using Gumbel) is not used because below we also consider independence between blocks.
  \item Pairwise independence using mutual information: Two 1D RVs $X$ and $Y$ are independent if and only if their mutual information vanishes,
  $I(X,Y) = H(X,Y)- H(X) - H(Y)=0$ \cite{Granger1994}. In our case, the marginals are $U(0,1)$ and $H(X)=H(Y)=0$, hence $I(X,Y) = H(X,Y)$.
  This suggests a statistical test for the hypothesis that $X$ and $Y$ are independent as follows.
  Suppose $X$ and $Y$ are independent. Draw $N$ independent samples and plot the density of the 2D entropy $H(X,Y)$.
  For a given acceptance threshold $\alpha$, find the cutoff value $H_{2,c}$ such that $P(H(X,Y)<H_{2,c})=1-\alpha$.
  Figure~\ref{fig:H2} shows the distribution for different values of $N$.
  With $\alpha=0.05$, the cutoff can be approximated by $H_{2,c}=-0.75 N^{0.62}$.
  Accordingly, any pair of marginals which were found to be statistically uncorrelated, are also tested for independence using they mutual information (see below).
  \item 2D entropy: Two-dimensional entropy (which, in our case, is always compact with support $[0,1]^2$) is estimated using a 2D histogram
  with uniformly spaced bins in each dimension.
\end{enumerate}

As a final note, we address the choice of which dimension should be used for splitting in the recursion step.
We suggest splitting the dimension which shows the strongest correlations with other marginals.
To this end, we square the elements in the correlation matrix $R$ and sum the rows.
We pick the column with the largest sum (or the first of them if several are equal).

Lastly, we consider the computational cost of the algorithm,
which has four components whose efficiency requires consideration:
\begin{enumerate}
\item Sorting of 1D samples: In the first level, samples may be unbounded and sorting can cost $O(N \log N)$.
However, for the next levels, the samples are approximately uniformly distributed in $[0,1]$ and bucket sort works with an average cost of $O(N)$. This is multiplied by the number of levels, which is $O(\log N)$.
As all $D$ dimensions need to be sorted, the overall cost of sorting is $O(D N \log N)$.
\item Calculating 1D entropies. Since the data is already sorted, calculating the entropy using either binning or spacing has a cost $O(N)$ per dimension, per level. Overall $O(DN\log N)$.
\item Pairwise correlations: $D(D-1)/2$ pre-sorted pairs, each costs $O(N)$ per level.
Overall $O(D^2 N \log N)$.
\item Pairwise entropy: The worst-case is that all pairs are uncorrelated but dependent, which implies that all pairwise mutual information need to be calculated at all levels. However, pre-sorting again reduces the cost of calculating histograms to $O(N)$ per level. 
With $O(\log N)$ levels, the cost is $O(D^2 N \log N)$.
\end{enumerate}
Overall, the cost of the algorithm is $O(D^2 N \log N)$.
The bottleneck is due to the stopping criterion for the recursion.
A simpler test may reduce the cost by a factor $D$.
However, in addition to the added accuracy, checking for pairwise independence allows, for some distributions, splitting the samples into several lower dimensional estimates which is both efficient and more accurate.

%%%%%%%%%%%%%%%%%%%%%%%%%%%%%%%%%%%%%%%%%%%%%%%%%%%%%%%%%
\section{Summary}
\label{sec:summary}
%%%%%%%%%%%%%%%%%%%%%%%%%%%%%%%%%%%%%%%%%%%%%%%%%%%%%%%%%

We presented a new algorithm for estimating the differential entropy of high-dimensional
distributions using independent samples.
The method applies the idea of decoupling the entropy to a sum of 1D contributions,
corresponding to the entropy of marginals, and the entropy of the copula, describing
the dependence between the variables.
Marginal densities are estimated using known methods for scalar distributions.
The entropy of the copula is estimated recursively, similar to the $k$-D partitioning
tree method.
Our numerical examples demonstrate the applicability of our method up to a dimension of 50,
showing improved accuracy and efficiency compared to previously suggested schemes.
The main disadvantage of the algorithm is the assumption that pair-wise independent components of the data are truly independent.
This approximations may clearly fail for particularly chosen setups.
Here, we focused on the presentation of the algorithm and demonstrated its advantages.
Rigorous proofs of consistency and analysis of convergence rates are beyond the scope of the present manuscript.

Our tests demonstrates that compression-based methods do not provide accurate estimates of the entropy, at least for the synthetic examples tested.
Nonetheless, it is surprising that some quantitative estimate of entropy can be obtained using such simple-to-implement method.
Moreover, this approach can be easily applied to high-dimensional distributions. 
Under some ergodic or mixing properties, independent sampling can be easily replaced by larger ensembles.
Thus, for dimension 100 or higher (e.g., a 50 particles system in 2D), all the direct estimation methods (kDP, kNN and CADEE) are 
prohibitively expensive.

To conclude, our numerical experiments suggest that $k$NN methods  are favorable for unbounded distributions up to about dimension 20.
At higher dimensions $k$NN becomes both inaccurate and highly inefficient. 
For distribution with compact support, or when the support is mixed or unknown, the proposed CADEE method is significantly more robust.
At relatively high dimensions (up to 100), the CADEE method provides
reliable estimates even under severe under-sampling and at a reasonable computational cost.
We suggest using the recursive copula splitting scheme
for other applications requiring estimation of copulas and evaluation of mutual dependencies between RVs, for example, in financial applications and neural signal processing algorithms.

A Matlab code is available at \url{arielg333.wixsite.com/personal/cadee} (a GitHub or similar version with examples is currently being developed).

\section*{Acknowledgments}
GA thanks partial support from The Israel Science Foundation Grant No. 373/16 and
The Deutsche Forschungsgemeinschaft (The German Research Foundation DFG) Grant No. BA1222/7-1.

\bibliography{bibfile}

\clearpage

\clearpage

\begin{table}[h!t]
\caption{Estimating the entropy for given analytically-computable examples at $D=10$. {\bf The best method is highlighted in bold.}}
\centering{}%
\begin{tabular}{|c|c|c|c|c|c|}
\hline
Example & exact & CADEE & $k$DP & $k$NN & compression \\
\hline
C1 - uniform & 0 & -1.5e-3 & {\bf -7.16e-4} & 0.81 & -4.3e-2 \\
C2 - pairs & -0.45 & {\bf -0.46} & -0.32 & 0.30 & -0.25 \\
C3 - boxes & -20.7 & {\bf -20.6} & -5.3 & -19.7 & 1.3e-4 \\
UB1 - Gauss & -0.915 & -1.3 & 9.1 & {\bf -0.9} & 5.1 \\
UB2 - power-law & 12.6 & 15.7 & 92.3 & {\bf 14.7} & 67.2 \\
\hline
\end{tabular}
\label{tbl:results10}
\end{table}

\vspace{2cm}

\begin{table}[h!t]
\caption{Estimating the entropy for given analytically-computable examples at $D=20$. {\bf The best method is highlighted in bold.}}
\centering{}%
\begin{tabular}{|c|c|c|c|c|c|}
\hline
Example & exact & CADEE & $k$DP & $k$NN & compression \\
\hline
C1 - uniform      & 0      & -2.9e-3 & {\bf -3.4e-4} & 3.3 & -1.1e-2 \\
C2 - pairs          & -0.91 & {\bf -0.98}   & -0.50    & 2.3 & -0.43 \\
C3 - boxes         & -56.9 & -60.6   & -5.15    & {\bf -52.9} & 5.8e-3 \\
UB1 - Gauss       & -14.0 & {\bf -14.4}   & 18.6     & -12.6 & 5.0 \\
UB2 - power-law & 30.2  & 47.2    & 296.6   & {\bf 40.3} & 131.6 \\
\hline
\end{tabular}
\label{tbl:results20}
\end{table}

\clearpage
\appendix

\setcounter{figure}{0} \renewcommand{\thefigure}{A.\arabic{figure}}

\section*{Additional pseudo-code used for numerical examples}

Multiple methods can be used for  estimating the 1D entropy, we applied the following pseudo-code.
 \begin{algorithmic}[H]
   \Function{H1D}{$\{ u^i \}$, $N$, $level$}
   \If{$level=0$}
   \State \Comment{$M_n$ Spacing method}
   \State $M_n \gets$ round($N^{1/3}$)
   \State $\Delta^i \gets u^{M_n+i} - u^i$, $i=1,\dots,N-M_n$
   \State $H \gets \left[ \sum_{i=1}^{M_n} \ln \Delta^i + (N-M_n) \ln(N/M_n) \right]/N$
   \Else
   \Comment{Uniform bins in $[0,1]$}
   \State $N_{\rm bins} \gets \min \{ {\rm max \# bins} , N^{0.4}, N/10 \}$
   \State $edges \gets [0 , 1/N_{\rm bins}, 2/N_{\rm bins}, \dots , 1]$
   \State  \Comment{Histogram with bins $edges$}
   \State $counts \gets$ Histogram ($\{ u^i \}$,$edges$)
   \State $p \gets counts \cdot N_{\rm bins}/(\sum counts)$ \Comment{Normalize}
   \State $H \gets -\left[ \sum p\ln(p) \right]/N_{\rm bins}$ \Comment{where $0 \ln 0=0$}
   \EndIf
   \State \Return $H$
   \EndFunction
 \end{algorithmic}

We suggest the following pseudo-code for estimating independence of two 1D RVs  (already the rank vectors), which was used in the numerical examples.
\begin{algorithmic}[H]
  \Function{areIndependent}{X,Y,$N$}
    \State $R=$corr($X$,$Y$)  \Comment{Pearson correlation}
    \State $I \gets $(P-value$(R,N)<\alpha$) \Comment{true if statistically uncorrelated}
    \If{$I$=false}
    \State $H_2 = $ H2D($X,Y$) \Comment{Calculate 2D entropy}
    \If{$H_2 N^{0.62}<-0.75$}
    \State $I \gets$ true
    \EndIf
    \EndIf
\EndFunction
\end{algorithmic}

   \begin{algorithmic}[H]
     \Function{H2D}{$X$,$Y$,$N$}
     \State \Comment{calc 2D histogram using uniform bins}
     \State $N_{\rm bins} \gets \min \{ {\rm max \# bins} , N^{0.2}, N/10 \}$
     \State $edges \gets [0 , 1/N_{\rm bins}, 2/N_{\rm bins}, \dots , 1]$
     \State \Comment{Histogram with bins $edges$ per dim}
     \State $counts \gets$ 2D Histogram($X$,$Y$,$edges$)
     \State $p \gets counts \cdot N_{\rm bins}^2/(\sum counts)$ \Comment{Normalize}
     \State $H \gets -\left[ \sum p\ln(p) \right]/N_{\rm bins}^2$
     \EndFunction
   \end{algorithmic}

\begin{figure}[h!tp]
\centering
\includegraphics[trim={0cm 0cm 0cm 0cm},clip, width=8cm, angle=0]{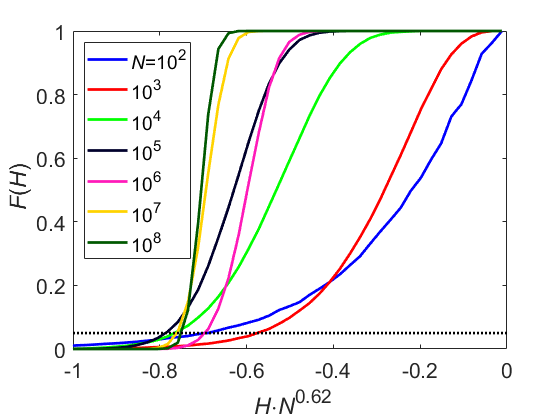}
\caption{
 Numerical evaluation of the cumulative distribution function for the entropy of two scalar,
 independent, uniformly distributed random variables.
 After scaling with the sample size, we find that $P(H N^{-0.62}<-0.75)$ is approximately 0.05.
 Hence, it can be considered as a statistics for accepting the hypothesis that the random variables are independent.
}
\label{fig:H2}
\end{figure}

\end{document}